\documentclass[aps,showkeys,twocolumn,preprintnumbers,amsmath,amssymb,superscriptaddress,floatfix,nofootinbib]{revtex4}
\usepackage{graphicx,color,dcolumn,booktabs,bm}
\usepackage{amsmath}
\usepackage{graphicx}
\usepackage{longtable,lscape}
\usepackage{txfonts}
\usepackage{overpic}
\usepackage{epsfig}
\usepackage{amssymb}
\usepackage{bbding}
\usepackage{rotating}
\usepackage{epstopdf}
\usepackage{appendix}
\usepackage{indentfirst}
\usepackage{feynmf}   
\usepackage{slashed}  
\usepackage{cases}
\usepackage{color}
\usepackage{multirow}
\usepackage{graphicx,color,dcolumn,booktabs,bm}
\usepackage{cases}
\usepackage{array}

\graphicspath{{Figures/}} %

\usepackage[colorlinks, citecolor=blue,anchorcolor=red,menucolor=red, linkcolor=red,filecolor=red,runcolor=red,urlcolor=blue,frenchlinks=red]{hyperref}

\begin{document}
\title{Predicting hidden bottom molecular tetraquarks with a complex scaling method}
\author{Qing-Fu Song}

\affiliation{Department of Physics, Hunan Normal University, Changsha 410081, China}
\affiliation{Key Laboratory of Low-Dimensional Quantum Structures and Quantum Control of Ministry of Education, Changsha 410081, China}
\affiliation{Key Laboratory for Matter Microstructure and Function of Hunan Province, Hunan Normal University, Changsha 410081, China}

\author{Qi-Fang L\"{u}}\email{lvqifang@hunnu.edu.cn}

\affiliation{Department of Physics, Hunan Normal University, Changsha 410081, China}
\affiliation{Key Laboratory of Low-Dimensional Quantum Structures and Quantum Control of Ministry of Education, Changsha 410081, China}
\affiliation{Key Laboratory for Matter Microstructure and Function of Hunan Province, Hunan Normal University, Changsha 410081, China}

\author{Dian-Yong Chen}\email{chendy@seu.edu.cn} %
\affiliation{School of Physics, Southeast University, Nanjing 210094, China}
\affiliation{
Lanzhou Center for Theoretical Physics, Lanzhou University, Lanzhou 730000, China}

\author{Yu-Bing Dong}\email{dongyb@ihep.ac.cn} %
\affiliation{Institute of High Energy Physics, Chinese Academy
of Sciences, Beijing 100049, China}
\affiliation{School of Physical Sciences, University of Chinese Academy of Sciences, Beijing 101408, China}

\begin{abstract}
In present work, we perform a coupled-channel analysis of $B^{(*)}_{(s)}\bar{B}^{(*)}_{(s)}$ systems with the one-boson-exchange potentials. We first study the $I(J^{PC})=1(1^{+-})$ $B\bar{B}^{*}/B^{*}\bar{B}^{*}$ system to describe the $Z_{b}(10610)$ and $Z_{b}(10650)$ particles as molecular states and determine the reasonable range of cutoff parameter $\Lambda$. Then, other $B^{(*)}_{(s)}\bar{B}^{(*)}_{(s)}$ combinations with different quantum numbers are systematically investigated. Some bound states and resonances appear in the isoscalar systems, while only several shallow bound states exist for isovector systems. Far away from the excited conventional $P-$wave bottomium, these predicted states can be easily identified as exotic particles both theoretically and experimentally. Moreover, the $\eta_b(nS)/\Upsilon(nS)$ plus light mesons are the excellent final states to search for the bound states, while the $B\bar B^*+h.c.$ and $B^* \bar B^*$ channels are suitable for the resonances. We highly recommend that the LHCb and Belle II Collaborations can hunt for these bottomonium-like states in future.
\end{abstract}


\maketitle
\section{introduction}
During the last two decades, many new hadron states have been observed in the large scientific facilities, which triggered plenty of theoretical interest and revitalized the study of hadronic spectroscopy~\cite{Hosaka:2016pey,Lebed:2016hpi,Chen:2016qju,Ali:2017jda,
Esposito:2016noz,Dong:2017gaw,Guo:2017jvc,Olsen:2017bmm,
Karliner:2017qhf,Liu:2019zoy,Brambilla:2019esw,Bicudo:2022cqi,Mai:2022eur,
Barabanov:2020jvn,Chen:2022asf}. Past researches show that some of the new hadrons are hard to classified into the conventional mesons or baryons. Then, new effective degrees of freedom need to be introduced and more complicated configurations named  exotic states emerge. Beyond the traditional quark-antiquark and three-quark pictures, quantum chromodynamics (QCD) actually allows the existence of plentiful hadronic structures, such as molecular states, compact multiquarks, hadro-quarkonium, hybrid states, glueball and their mixtures. Since many new particles lie near the hadron-hadron thresholds, the picture of hadronic molecule is significantly competitive among these theoretical interpretations~\cite{Chen:2022asf,Dong:2017gaw,Guo:2017jvc,Karliner:2017qhf,Zou:2021sha,Mai:2022eur,Zou:2013af}. Thus, it becomes a primary task to pick out the genuine molecular states from lots of candidates and predict more ones for experimental searches.

\begin{figure}
    \centering
    \includegraphics[scale=0.68]{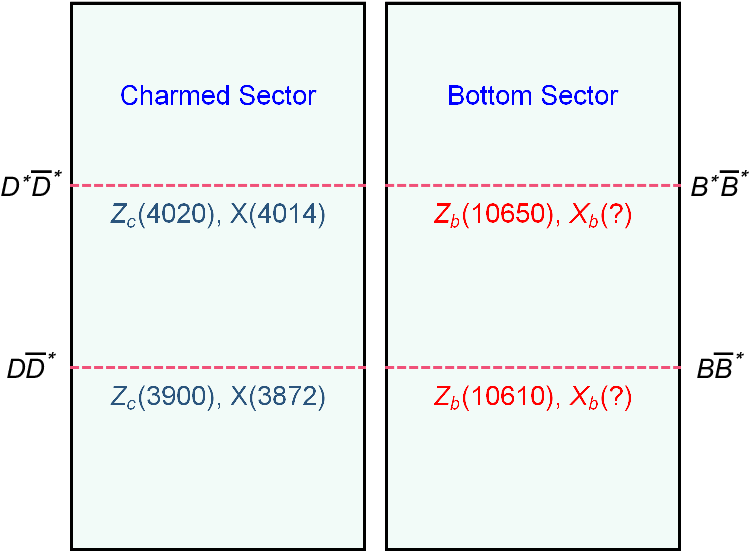}
    \caption{Comparison of  exotic states between hidden charm and bottom sectors.}
    \label{spectrum}
\end{figure}

Among lots of molecular candidates, the states composed of ground state hadrons attract great attentions both experimentally and theoretically.  The longstanding charmonium-like state $X(3872)$  was first observed in the process $B^{\pm}$$\to $$K^{\pm}\pi^{+}\pi^{-}J/\psi$  by the Belle Collaboration in 2003~\cite{Belle:2003nnu}, which opened a new era in the investigations of exotic states. Since the $X(3872)$ locates very near the $D \bar D^{*} +h.c.$ threshold, which could be naturally interpreted as a  hadronic  molecular state.  In 2013, a charged  charmonium-like state named $Z_{c}(3900)$ is reported by the BESIII~\cite{BESIII:2013ris} and Belle~\cite{Belle:2013yex} Collaborations, and arouses more attentions. The $Z_{c}(3900)$ lies slightly above the $D \bar D^{*} +h.c.$ threshold, which can be regarded as the isospin partner of $X(3872)$. According to  heavy quark spin symmetry, in the corresponding $D^*\bar{D}^{*}$ threshold, there should also be similar charmonium-like states. Indeed, an isovector state $Z_c(4020)$  and an isoscalar state $X(4013)$ were subsequently discovered by the BESIII~\cite{BESIII:2013ouc} and Belle~\cite{Belle:2021nuv} Collaborations, respectively. However, in the bottom sector, only two bottomonium-like states with $I(J^{PC})=1(1^{+-})$ were reported by the Belle Collaboration~\cite{Belle:2011aa}, which are $Z_{b}(10610)$ with a mass of $10607.2^{+2}_{-2}$ MeV and a width $18.4^{+2.4}_{-2.4}$ MeV, and $Z_{b}(10650)$ with a mass of $10652.2^{+1.5}_{-1.5}$ MeV and a width $11.5^{+2.2}_{-2.2}$ MeV, respectively \footnote{Hereafter, when we discuss the $C$-parity of the isospin vector states, we refer to the one of the neutral state.}. The comparison of exotic states between hidden charm and bottom sectors are displayed in Figure~\ref{spectrum}. It is clear that some undiscovered bottomonium-like states near the $B \bar B^{*} +h.c.$ and $B^* \bar B^{*}$ thresholds are expected to complete this regular pattern. 

Theoretically, the bottomonium-like states $Z_b(10610)$ and $Z_b(10650)$ were widely investigated in the literature. Most of works suggest that these two states can be explained as the $B \bar B^{*} +h.c.$ and $B^* \bar B^{*}$ molecules, respectively~\cite{Bondar:2011ev,Sun:2011uh,Mehen:2011yh,Cleven:2011gp,Dias:2014pva,Li:2012wf,Yang:2011rp,Zhang:2011jja,Wang:2013daa, Wang:2014gwa,Dong:2012hc,Ohkoda:2013cea,Li:2012as,Cleven:2013sq,Li:2012uc,Li:2014pfa,Xiao:2017uve,Wu:2018xaa,Wu:2020edh,Voloshin:2011qa,Cheng:2023vyv}, though other interpretations such as compact tetraquarks and kinematical effects are possible~\cite{Guo:2011gu,Cui:2011fj,Wang:2013zra,Chen:2011pv,Chen:2012yr}. Also, it is easy to see that the mass splittings $M[Z_b(10650)]-M[Z_b(10610)]=M(B^*)-M(B)$ and $M[Z_c(4020)]-M[Z_c(3900)]=M(D^*)-M(D)$ satisfy the heavy quark symmetry well, which strongly indicates that the $Z_b(10610)$ and $Z_b(10650)$ states are bottom analogues of $Z_c(3900)$ and $Z_c(4020)$. Under the molecular scenario, all the resonances shown in Figure~\ref{spectrum} can be uniformly understood, and other promising $X_b$ states call for further exploration. 

In previous molecular calculations for $Z_b(10610)$ and $Z_b(10650)$, one usually solved the single channel stationary Schr\"odinger equation in one-boson-exchange model to obtain weekly bound molecular states. However, the coupled-channel effects among the whole hidden bottom channels may contribute significantly for certain resonances. Moreover, the standard stationary Schr\"odinger equation can only handle with the weekly bound molecular states, while a genuine physical state may show itself in other ways, such as a resonance above the threshold. Although the Belle collaboration declaimed that the $Z_b(10610)$ and $Z_b(10650)$ lie above the $B\bar{B}^*$ and $B^*\bar{B}^*$ thresholds, the precise properties of these near threshold states may depend on the schemes of parametrization as indicated in Ref.~\cite{Hanhart:2015cua} and there also exist some theoretical works supporting that these two exotic states should lie below the corresponding thresholds \cite{Sun:2011uh, He:2024aej, Dias:2014pva, Zhao:2021cvg}. We hold the opinion that whether the $Z_b(10610)$ and $Z_b(10650)$ locating above or below the corresponding thresholds is still an open question and worthy of further study. Thus, it is urgent to study the $B^{(*)}_{(s)}\bar{B}^{(*)}_{(s)}$ systems systematically in a coupled-channel approach and look for possible existing resonances together with bound states.

In present work, we perform a coupled-channel analysis of $B^{(*)}_{(s)}\bar{B}^{(*)}_{(s)}$ systems. With the one-boson-exchange model, the coupled-channel Schr\"odinger equation are solved by using the Guassian expansion method~\cite{Hiyama:2003cu,Hiyama:2018ivm} and complex scaling method  (CSM)~\cite{Aguilar:1971ve,Balslev:1971vb,Moiseyev:1998gjp, Ho:1983lwa}. We first calculate the $Z_{b}(10610)$ and  $Z_{b}(10650)$ in the coupled $B\bar{B}^{*}+h.c.$ and $B^{*}\bar{B}^{*}$  systems with $I(J^{PC})=1(1^{+-})$, which support the molecular pictures in a reasonable range for the cutoff parameter $\Lambda$. Then, we systematically study various $B^{(*)}_{(s)}\bar{B}^{(*)}_{(s)}$ combinations with different quantum numbers to search for possible bound states and resonances. We predict some bottomonium-like states with $I(J^P)=0(1^{++})$ $X_b$ particles lie near the $B\bar{B}$, $B\bar{B}^{*}+h.c.$, respectivelt, and $B^{*}\bar{B}^{*}$ thresholds, which urgently need to search for by future experiments.

This paper is organized as follows. The formalism of effective interactions and complex scaling method are briefly introduced in Sec.~\ref{model}. We present the numerical results and discussions for the $B^{(*)}_{(s)}\bar{B}^{(*)}_{(s)}$ systems in Sec.~\ref{results}. A summary is given in the last section.

\section{Formalism}\label{model}
\subsection{The effective interactions}
The one-boson-exchange model is successful in analyzing the formation mechanisms of hadronic molecules, and here it is applied to study the  $B^{(*)}_{(s)}\bar{B}^{(*)}_{(s)}$ systems. Satisfying the heavy quark symmetry and chiral symmetry, the relevant effective Lagrangians can be expressed as~\cite{Yan:1992gz,Wise:1992hn,Burdman:1992gh,Casalbuoni:1996pg},
\begin{eqnarray}
{\mathcal L}&=&g_{\sigma}\left\langle H^{(Q)}_a\sigma\overline{H}^{(Q)}_a\right\rangle+g_{\sigma}\left\langle \overline{H}^{(\bar{Q})}_a\sigma H^{(\bar{Q})}_a\right\rangle\nonumber\\
&&+ig\left\langle H^{(Q)}_b{\mathcal A}\!\!\!\slash_{ba}\gamma_5\overline{H}^{\,({Q})}_a\right\rangle+ig\left\langle \overline{H}^{(\bar{Q})}_a{\mathcal A}\!\!\!\slash_{ab}\gamma_5 H^{\,(\bar{Q})}_b\right\rangle\nonumber\\
&&+\left\langle iH^{(Q)}_b\left(\beta v^{\mu}({\mathcal V}_{\mu}-\rho_{\mu})+\lambda \sigma^{\mu\nu}F_{\mu\nu}(\rho)\right)_{ba}\overline{H}^{\,(Q)}_a\right\rangle\nonumber\\
&&-\left\langle i\overline{H}^{(\bar{Q})}_a\left(\beta v^{\mu}({\mathcal V}_{\mu}-\rho_{\mu})-\lambda \sigma^{\mu\nu}F_{\mu\nu}(\rho)\right)_{ab}H^{\,(\bar{Q})}_b\right\rangle,
\end{eqnarray}
where $a$ and $b$ is the flavor indices, and $v^{\mu}=(1, \bf{0})$ is the four-velocity of the heavy hadron. The vector current, axial current, and strength tensor of vector field are defined as,
\begin{eqnarray}
    \mathcal{V}_{\mu} &= &\frac{1}{2}(\xi^{\dag}\partial_{\mu}\xi+\xi\partial_{\mu}\xi^{\dag})\\
A_{\mu} &= &\frac{1}{2}(\xi^{\dag}\partial_{\mu}\xi-\xi\partial_{\mu}\xi^{\dag})\\
F_{\mu\nu}(\rho)&=&\partial_{\mu}\rho_{\nu}-\partial_{\nu}\rho_{\mu}+[\rho_{\mu},\rho_{\nu}]
\end{eqnarray}
with $\xi=\text{exp}(i{\mathbb{P}}/f_{\pi})$ and $\rho_{\mu}=ig_{V}{\mathcal{V}}_{\mu}/\sqrt{2}$. The $\mathbb{P}$ and $\mathbb{V}$ refer to the matrices of light pseudoscalar and vector mesons, respectively,  which are,
\begin{eqnarray}
\left.\begin{array}{c} {\mathbb{P}} = {\left(\begin{array}{ccc}
       \frac{\pi^0}{\sqrt{2}}+\frac{\eta}{\sqrt{6}} &\pi^+ &K^+\\
       \pi^-       &-\frac{\pi^0}{\sqrt{2}}+\frac{\eta}{\sqrt{6}} &K^0\\
       K^-         &\bar K^0   &-\sqrt{\frac{2}{3}} \eta     \end{array}\right)},\\
{\mathbb{V}} = {\left(\begin{array}{ccc}
       \frac{\rho^0}{\sqrt{2}}+\frac{\omega}{\sqrt{2}} &\rho^+ &K^{*+}\\ 
       \rho^-       &-\frac{\rho^0}{\sqrt{2}}+\frac{\omega}{\sqrt{2}} &K^{*0}\\
       K^{*-}         &\bar K^{*0}   & \phi     \end{array}\right)}.
\end{array}\right.
\end{eqnarray}
The $H^{(Q)}_a$, $H^{(\bar Q)}_a$, $\bar H^{(Q)}_a$, and $\bar H^{(\bar Q)}_a$ represent the fields of $S$-wave heavy-light mesons with the definitions  
\begin{eqnarray}
H^{(Q)}_a&=&(1+{v}\!\!\!\slash)(\mathcal{P}^{*(Q)\mu}_a\gamma_{\mu}-\mathcal{P}^{(Q)}_a\gamma_5)/2, \nonumber\\
H^{(\bar{Q})}_a &=& (\bar{\mathcal{P}}^{*(\bar{Q})\mu}_a\gamma_{\mu}-\bar{\mathcal{P}}^{(\bar{Q})}_a\gamma_5)
(1-{v}\!\!\!\slash)/{2},\nonumber\\
\bar{H}&=&\gamma_0H^{\dagger}\gamma_0.
\end{eqnarray}
More explicitly, the effective Lagrangian depicting the couplings of light mesons and heavy-light mesons can be written as~\cite{Chen:2022dad,Li:2012cs,Sun:2011uh,Liu:2017mrh}
\begin{eqnarray}
	\mathcal{L}_{\mathcal{P} \mathcal{P} \mathbb{V}}&= & -\sqrt{2} \beta g_{V} \mathcal{P}_{b} \mathcal{P}_{a}^{\dagger} v \cdot \mathbb{V}_{b a}+\sqrt{2} \beta g_{V} \widetilde{\mathcal{P}}_{a}^{\dagger} \widetilde{\mathcal{P}}_{b} v \cdot \mathbb{V}_{a b},\nonumber\\
	\mathcal{L}_{\mathcal{P}^{*} \mathcal{P} \mathbb{V}}&= & -2 \sqrt{2} \lambda g_{V} v^{\lambda} \varepsilon_{\lambda \mu \alpha \beta}\left(\mathcal{P}_{b} \mathcal{P}_{a}^{* \mu \dagger}+\mathcal{P}_{b}^{* \mu} \mathcal{P}_{a}^{\dagger}\right)\left(\partial^{\alpha} \mathbb{V}^{\beta}\right)_{b a} \nonumber\\
	& -&2 \sqrt{2} \lambda g_{V} v^{\lambda} \varepsilon_{\lambda \mu \alpha \beta}\left(\widetilde{\mathcal{P}}_{a}^{* \mu \dagger} \widetilde{\mathcal{P}}_{b}+\widetilde{\mathcal{P}}_{a}^{\dagger} \widetilde{\mathcal{P}}_{b}^{* \mu}\right)\left(\partial^{\alpha} \mathbb{V}^{\beta}\right)_{a b},\nonumber\\
	\mathcal{L}_{\mathcal{P}^{*} \mathcal{P}^{*} \mathbb{V}}&= & \sqrt{2} \beta g_{V} \mathcal{P}_{b}^{*} \cdot \mathcal{P}_{a}^{* \dagger} v \cdot \mathbb{V}_{b a} \nonumber\\
	& -&i 2 \sqrt{2} \lambda g_{V} \mathcal{P}_{b}^{* \mu} \mathcal{P}_{a}^{* v^{\dagger}}\left(\partial_{\mu} \mathbb{V}_{v}-\partial_{v} \mathbb{V}_{\mu}\right)_{b a} \nonumber\\
	& -&\sqrt{2} \beta g_{V} \widetilde{\mathcal{P}}_{a}^{* \dagger} \widetilde{\mathcal{P}}_{b}^{*} v \cdot \mathbb{V}_{a b} \nonumber\\
	& -&i 2 \sqrt{2} \lambda g_{V} \widetilde{\mathcal{P}}_{a}^{* \mu \dagger} \widetilde{\mathcal{P}}_{b}^{* v}\left(\partial_{\mu} \mathbb{V}_{v}-\partial_{v} \mathbb{V}_{\mu}\right)_{a b},\nonumber\\
	\mathcal{L}_{ \mathcal{P} \mathcal{P{\sigma}}} & =&-2 g_{s} \mathcal{P}_{b} \mathcal{P}_{b}^{\dagger} \sigma-2 g_{s} \widetilde{\mathcal{P}}_{b} \widetilde{\mathcal{P}}_{b}^{\dagger} \sigma, \nonumber\\
	\mathcal{L}_{\mathcal{P}^{*} \mathcal{P}^{*} \sigma} & =&2 g_{s} \mathcal{P}_{b}^{*} \cdot \mathcal{P}_{b}^{* \dagger} \sigma+2 g_{s} \widetilde{\mathcal{P}}_{b}^{*} \cdot \widetilde{\mathcal{P}}_{b}^{* \dagger} \sigma.
\end{eqnarray}

With the above Lagrangians, the relevant potentials can be derived straightforwardly. Based on the Breit approximation, the effective potential in momentum space reads
\begin{eqnarray}\label{breit}
\mathcal{V}^{h_{1}h_{2}\to h_{3}h_{4}}(\bm{q}) &=&
          -\frac{\mathcal{M}(h_{1}h_{2}\to h_{3}h_{4})}
          {4\sqrt{m_{1}m_{2}m_{3}m_{4}}},
\end{eqnarray}
where $\mathcal{M}(h_{1}h_{2}\to h_{3}h_{4})$ denotes the scattering amplitude for the $h_{1}h_{2}\to h_{3}h_{4}$ process, and $m_{i}$ is the mass of the particle $h_i$. By performing the Fourier transformation, one can obtain the final effective potential in position space
\begin{eqnarray}
\mathcal{V}(\bm{r}) =
          \int\frac{d^3\bm{q}}{(2\pi)^3}e^{i\bm{q}\cdot\bm{r}}
          \mathcal{V}(\bm{q})\mathcal{F}^2(q^2,m_i^2).
\end{eqnarray}
The form factor $\mathcal{F}(q^2,m_i^2)= (\Lambda^2-m_i^2)/(\Lambda^2-q^2)$ with the cutoff parameter $\Lambda$ is introduced to account for the inner structures of the interacting hadrons.

Then, we can obtain the flavor-independent sub-potentials
\begin{eqnarray}
V^{a}_{v}&=& -\frac{1}{2}\beta^{2}g_{V}^{2}Y(\Lambda,m_{v},r)\nonumber,\\
V^{a}_{\sigma}&=&-g_{s}^{2}Y(\Lambda,m_{\sigma},r),
\end{eqnarray}
for $PP\to PP$ process,
\begin{eqnarray}
V^{c}_{p}&=&-\frac{1}{3}\frac{g^{2}}{f_{\pi}^{2}} \mathcal{Z}_{\Lambda, m_{p}},\nonumber\\
V^{c}_{v}&=&\frac{2}{3}\lambda^{2}g_{V}^{2}\mathcal{Z}_{\Lambda, m_{v}}^{\prime},
\end{eqnarray}
for $PP\to P^{*}P^{*}$ process,
\begin{eqnarray}
    V^{d}_{\sigma}&=&-g_{s}^{2}\mathcal{Y}_{\Lambda, m_{\sigma}},\nonumber\\
    V^{d}_{v}&=&-\frac{1}{2}\beta^{2}g_{V}^{2}\mathcal{Y}_{\Lambda, m_{v}},
\end{eqnarray}
for $ P^{}P^{*}\to P^{}P^{*}$ process, 
    \begin{eqnarray}
V^{e}_{p}&=&-\frac{1}{3}\frac{g^{2}}{f_{\pi}^{2}} \mathcal{Z}_{\Lambda, m_{p}},\nonumber\\
V^{e}_{v}&=&\frac{2}{3}\lambda^{2}g_{V}^{2}\mathcal{Z}_{\Lambda, m_{v}}^{\prime},
\end{eqnarray}
for $ P^{}P^{*}\to P^{*}P^{}$ process, and
\begin{eqnarray}\nonumber
V_{p}^{f} &=& -\frac{1}{3}\frac{g^2}{f_{\pi}^2}\mathcal{Z}_{\Lambda_{i},m_{i}}^{ij},\nonumber\\
    V_{v}^{f} &=&\frac{2}{3}\lambda^2g_V^2
    \mathcal{X}_{\Lambda_{i},m_{i}}^{ij},\nonumber\\
\end{eqnarray}
for $PP^{*}\to P^{*}P^{*}$ process. 
\begin{eqnarray}
       V_{p}^{g} &=& -\frac{1}{3}\frac{g^2}{f_{\pi}^2}\mathcal{Z}_{\Lambda_{i},m_{i}}^{ji},\nonumber\\
    V_{v}^{g} &=&\frac{2}{3}\lambda^2g_V^2\mathcal{X}_{\Lambda_{i},m_{i}}^{ji},
\end{eqnarray}
for $P^{*}P\to P^{*}P^{*}$ process, \begin{eqnarray}
V^{h}_{p}&=&-\frac{1}{3}\frac{g^{2}}{f_{\pi}^{2}} \mathcal{Z}_{\Lambda, m_{p}},\nonumber\\
V^{h}_{\sigma}&=&-g_{s}^{2}\mathcal{Y}_{\Lambda, m_{\sigma}},\nonumber\\
V^{h}_{v}&=&-\frac{1}{2}\beta^{2}g_{V}^{2}\mathcal{Y}_{\Lambda, m_{v}}+\frac{2}{3}\lambda^{2}g_{V}^{2}\mathcal{Z}_{\Lambda, m_{v}}^{\prime}.
\end{eqnarray}
for $ P^{*}P^{*}\to P^{*}P^{*}$ process. Some explicit formulas for the potentials can be expressed as
\begin{eqnarray}
\mathcal{Z}_{\Lambda,m_a}^{ij}&=&\Bigg(\mathcal{E}_{ij}\nabla^{2}+\mathcal{F}_{ij}r\frac{\partial}{\partial r}\frac{1}{r}\frac{\partial}{\partial r}\Bigg) Y(\Lambda,m,r),\nonumber\\\
\mathcal{Z}_{\Lambda,m_a}^{\prime ij}&=&\Bigg(2\mathcal{E}_{ij}\nabla^{2}-\mathcal{F}_{ij}r\frac{\partial}{\partial r}\frac{1}{r}\frac{\partial}{\partial r}\Bigg) Y(\Lambda,m,r),\nonumber\\\
\mathcal{X}^{ij}_{\Lambda,m_a}&=&\Bigg(-2\mathcal{E}_{ij}\nabla^{2}-(\mathcal{F}^{\prime}_{ij}-\mathcal{F}^{\prime\prime}_{ij})r\frac{\partial}{\partial r}\frac{1}{r}\frac{\partial}{\partial r}\Bigg) Y(\Lambda,m,r),\nonumber\\\
\mathcal{Y}^{ij}_{\Lambda,m_a}&=&\mathcal{D}_{ij} Y(\Lambda,m,r),\nonumber\\\
Y(\Lambda,m,r)&=&\frac{1}{4\pi r}(e^{-mr}-e^{-\Lambda r})-\frac{\Lambda^2-m^2}{8\pi\Lambda}e^{-\Lambda r},
\end{eqnarray}
where $\mathcal{D}_{ij}$, $\mathcal{E}_{ij}$, and $\mathcal{F}_{ij}$ denote as the operators for spin-spin coupling and  the tensor forces and relate with polarization vector  $\epsilon_i$ and Pauli matrix $\sigma$, respectively. For instance, $\mathcal{D}_{11}=\mathcal{E}_{11}=(\epsilon_2\cdot\epsilon^\dagger_4)$, $\mathcal{F}_{11}=S(\hat{r},\epsilon_2,\epsilon^\dagger_4)$ for $B\bar{B}^*\to B\bar{B}^*$ process, $\mathcal{E}_{12}=i\epsilon_3^\dagger\cdot(\epsilon^\dagger_4\times\epsilon_2)$, $\mathcal{F}_{12}=iS(\hat{r},\epsilon_3,\epsilon^\dagger_4\times\epsilon_2)$  for $B\bar{B}^*\to B^*\bar{B}^*$ process, and $\mathcal{D}_{22}=(\epsilon_1\cdot\epsilon_3^\dagger)(\epsilon_2\cdot \epsilon^\dagger_4),\mathcal{E}_{22}=(\epsilon_1\times\epsilon_3^\dagger)\cdot(\epsilon_2\times\epsilon^\dagger_4)$, $\mathcal{F}_{22}=S(\hat{r},\epsilon_1\times\epsilon_3^\dagger,\epsilon_2\times\epsilon^\dagger_4)$  for $B^*\bar{B}^*\to B^*\bar{B}^*$ process, where $S(\hat{r},\mathbf{a},\mathbf{b})=3(\hat{r}\cdot\mathbf{a})(\hat{r}\cdot\mathbf{b})-\mathbf{a}\cdot\mathbf{b}$.
The total effective potentials for $B^{(*)}_{(s)}\bar{B}^{(*)}_{(s)}$ systems can be expressed by the combinations of these sub-potentials~\cite{Chen:2022dad,Li:2012cs,Sun:2011uh,Liu:2017mrh} and summarized in Table~\ref{total}. Also, the relevant parameters are collected in Table~\ref{parameters}~\cite{Bando:1987br,Wang:2021yld,Chen:2022dad,Wang:2019aoc,Machleidt:1987hj}.

\begin{table*}[!htbp]
	\renewcommand\arraystretch{1.4}
	\caption{\label{total} The effective  potentials for $B^{(*)}_{(s)}\bar{B}^{(*)}_{(s)}$ systems.}
	\begin{ruledtabular}
		\begin{tabular}{cccccccc}
			&$I(J^{PC})=0(0^{++})$&$B\bar{B}$
			&$B_s\bar{B}_s$
			&$B^*\bar{B}^*$
			&$B_s^*\bar{B}_s^*$
			\\
			&$B\bar{B}$&$V^{a}_{\sigma}+\frac{3}{2}V_{\rho}^{a}+\frac{1}{2}V_{\omega}^{a}$&$\sqrt{2}V_{K^{*}}^{a}$& $\frac{1}{6}V^{b}_{\eta}+\frac{3}{2}V^{b}_{\pi}+\frac{3}{2}V_{\rho}^{b}+\frac{1}{2}V_{\omega}^{b}$&$\sqrt{2}V_{K^{*}}^{b}+\sqrt{2}V_{K}^{b}$\\
			&$B_s\bar{B}_s$&&$V^{a}_{\sigma}+V^{a}_{\phi}$&$\sqrt{2}V_{K^{*}}^{b}+\sqrt{2}V_{K}^{b}$&$\frac{2}{3}V_{\eta}^{b}+V_{\phi}^{b}$\\
			&$B^*\bar{B}^*$ &&&$V_{\sigma}^{c}+\frac{3}{2}V_{\pi}^{c}+\frac{1}{6}V_{\eta}^{c}+\frac{3}{2}V_{\rho}^{c}+\frac{1}{2}V_{\omega}^{c}$&$\sqrt{2}V_{K^{*}}^{c}+\sqrt{2}V_{K}^{c}$\\
			&$B_s^*\bar{B}_s^*$ &&&&$V_{\sigma}^{c}+\frac{2}{3}V_{\eta}^{c}+V_{\phi}^{c}$ \\
			\hline
			
			&$I(J^{PC})=0(1^{+\pm})$&$B\bar{B}^{*}$
			&$B^*\bar{B}^*$
			&$B_s\bar{B}_s^{*}$
			&$B_s^*\bar{B}_s^*$
			\\
			&\multirow{2}{*}{$B\bar{B}^{*}$}&$V^{d}_{\sigma}+\frac{3}{2}V_{\rho}^{d}+\frac{1}{2}V_{\omega}^{d}$&$\sqrt{\frac{1}{2}}(\frac{3}{2}V_{\pi}^{f}+\frac{1}{6}V_{\eta}^{f}+\frac{3}{2}V_{\rho}^{f}+\frac{1}{2}V_{\omega}^{f})$&$(\sqrt{2}V_{K^{*}}^{d})$&$\sqrt{\frac{1}{2}}(\sqrt{2}V_{K^{*}}^{f}+\sqrt{2}V_{K}^{f})$\\
			&&$\mp(\frac{3}{2}V^{e}_{\pi}+\frac{1}{6}V_{\eta}^{e}+\frac{3}{2}V^{e}_{\rho}+\frac{1}{2}V^{e}_{\omega})$&$\mp\sqrt{\frac{1}{2}}(\frac{3}{2}V_{\pi}^{g}+\frac{1}{6}V_{\eta}^{g}+\frac{3}{2}V_{\rho}^{g}+\frac{1}{2}V_{\omega}^{g})$&$\mp(\sqrt{2}V_{K^{*}}^{e}+\sqrt{2}V_{K}^{e})$&$\mp\sqrt{\frac{1}{2}}(\sqrt{2}V_{K^{*}}^{g}+\sqrt{2}V_{K}^{g})$\\
			&\multirow{2}{*}{$B^{*}\bar{B}^{*}$}&&\multirow{2}{*}{$V_{\sigma}^{c}+\frac{3}{2}V_{\pi}^{c}+\frac{1}{6}V_{\eta}^{c}+\frac{3}{2}V_{\rho}^{c}+\frac{1}{2}V_{\omega}^{c}$}&$\sqrt{\frac{1}{2}}(\sqrt{2}V_{K^{*}}^{f}+\sqrt{2}V_{K}^{f})$&\multirow{2}{*}{$\sqrt{2}V_{K^{*}}^{c}+\sqrt{2}V_{K}^{c}$}\\
			&&&&$\mp\sqrt{\frac{1}{2}}(\sqrt{2}V_{K^{*}}^{g}+\sqrt{2}V_{K}^{g})$&\\
			&\multirow{2}{*}{$B_s\bar{B}_s^{*}$ }&&&\multirow{2}{*}{$V^{d}_\sigma+V_{\phi}^d+\frac{2}{3}V_{\eta}^{e}+V_{\phi}$}&$\sqrt{\frac{1}{2}}(\frac{2}{3}V^{f}_{\eta}+V^{f}_{\phi})$\\
			&&&&&+$\sqrt{\frac{1}{2}}(\frac{2}{3}V^{g}_{\eta}+V^{g}_{\phi})$\\
			&$B_s^*\bar{B}_s^*$ &&&&$V_{\sigma}^{c}+\frac{2}{3}V_{\eta}^{c}+V_{\phi}^{c}$ \\\hline
			
			&$I(J^{PC})=0(2^{++})$
			&$B^*\bar{B}^*$
			&$B_s^*\bar{B}_s^*$\\
			&$B^*\bar{B}^*$& $V_{\sigma}^{c}+\frac{3}{2}V_{\pi}^{c}+\frac{1}{6}V_{\eta}^{c}+\frac{3}{2}V_{\rho}^{c}+\frac{1}{2}V_{\omega}^{c}$&$\sqrt{2}V_{K^{*}}^{c}+\sqrt{2}V_{K}^{c}$\\
			&$B_s^*\bar{B}_s^*$& &$V_{\sigma}^{c}+\frac{2}{3}V_{\eta}^{c}+V_{\phi}^{c}$ \\
			\hline
			
			&$I(J^{PC})=1(0^{++})$&$B\bar{B}$
			&$B^*\bar{B}^*$
			\\
			&$B\bar{B}$&$V^{d}_{\sigma}-\frac{1}{2}V_{\rho}^{d}+\frac{1}{2}V_{\omega}^{d}$& $\frac{1}{6}V^{b}_{\eta}-\frac{1}{2}-\frac{1}{2}V_{\rho}^{b}+\frac{1}{2}V_{\omega}^{b}$\\
			&$B^*\bar{B}^*$ &&$V_{\sigma}^{c}-\frac{1}{2}V_{\pi}^{c}+\frac{1}{6}V_{\eta}^{c}-\frac{1}{2}V_{\rho}^{c}+\frac{1}{2}V_{\omega}^{c}$\\\hline
			
				&$I(J^{PC})=1(1^{+\pm})$&$B\bar{B}^{*}$
			&$B^*\bar{B}^*$\\
			&\multirow{2}{*}{$B\bar{B}^{*}$}&$V^{d}_{\sigma}-\frac{1}{2}V_{\rho}^{d}+\frac{1}{2}V_{\omega}^{d}$&$\sqrt{\frac{1}{2}}(-\frac{1}{2}V_{\pi}^{f}+\frac{1}{6}V_{\eta}^{f}-\frac{1}{2}V_{\rho}^{f}+\frac{1}{2}V_{\omega}^{f})$\\
			&&$\mp(-\frac{1}{2}V^{e}_{\pi}+\frac{1}{6}V_{\eta}^{e}-\frac{1}{2}V^{e}_{\rho}+\frac{1}{2}V^{e}_{\omega})$&$\mp\sqrt{\frac{1}{2}}(-\frac{1}{2}V_{\pi}^{g}+\frac{1}{6}V_{\eta}^{g}-\frac{1}{2}V_{\rho}^{g}+\frac{1}{2}V_{\omega}^{g})$\\
			&$B^{*}\bar{B}^{*}$&&$V_{\sigma}^{c}-\frac{1}{2}V_{\pi}^{c}+\frac{1}{6}V_{\eta}^{c}-\frac{1}{2}V_{\rho}^{c}+\frac{1}{2}V_{\omega}^{c}$\\\hline

			&$I(J^{PC})=1(2^{++})$
			&$B^*\bar{B}^*$
			\\
			&$B^*\bar{B}^*$& $V_{\sigma}^{c}-\frac{1}{2}V_{\pi}^{c}+\frac{1}{6}V_{\eta}^{c}-\frac{1}{2}V_{\rho}^{c}+\frac{1}{2}V_{\omega}^{c}$\\
		\end{tabular}
	\end{ruledtabular}
\end{table*}

\begin{table}[htbp]
\caption{\label{parameters} The relevant parameters adopted in this work.}
\begin{ruledtabular}
\begin{tabular}{ccccccccc}
&Parameters
&Value
\\\hline
&$g_{V}$&5.800\\
&$g$&0.590\\
&$g_s$&2.820\\
&$\beta$&0.900\\
&$f_{\pi}$$(\textrm{GeV})$&0.132\\
&$\lambda$$(\textrm{GeV}^{-1})$&0.560\\
\end{tabular}
\end{ruledtabular}
\end{table}

\subsection{Complex scaling method}\label{sec3}
Here, we give a brief introduction for the complex scaling method. In the CSM, the relative distance $\bm{r}$ and its conjugate momentum $\bm{p}$ are transformed into the complex plane
by a transformation operator $U(\theta)$ with a real positive  scaling angle $\theta$, that is 
\begin{equation}
 \bm{r^{\prime}}=U(\theta)\bm{r}U^{-1}(\theta)=\bm{r}e^{i\theta}, \bm{p^{\prime}}=U(\theta)\bm{p}U^{-1}(\theta)=\bm{p}e^{-i\theta}
\end{equation}
with $U(\theta)U^{-1}(\theta)=1$. Then, the Schr\"odinger equation can be transformed as 
\begin{eqnarray}
H^{\theta}\Psi^{\theta}&=&E^{\theta}\Psi^{\theta}, H^{\theta}=U(\theta)HU^{-1}(\theta).
\end{eqnarray}
Consequently, the complex scaling Schr\"odinger
equation for the coupled channels can be written as
\begin{eqnarray}\label{S eq}
&&\left[\frac{1}{2\mu_{j}}\left(-\frac{d^{2}}{dr^{2}}+\frac{l_{j}(l_{j}+1)}{r^{2}}\right)e^{-2i\theta}+W_{j}\right]\psi_{j}^{\theta}(r) \nonumber \\ && \qquad +\sum _{k}V_{jk}(re^{i\theta})\psi_{k}^{\theta}(r) =E\psi_{j}^{\theta}(r),
\end{eqnarray}
where  $j$, $\mu_{j}$, $W_{j}$, and $\psi^{\theta}_j(r)$  represent the channel index, reduced mass, corresponding threshold, and the orbital wave function, respectively. 
\begin{figure}
    \centering
    \includegraphics[scale=0.65]{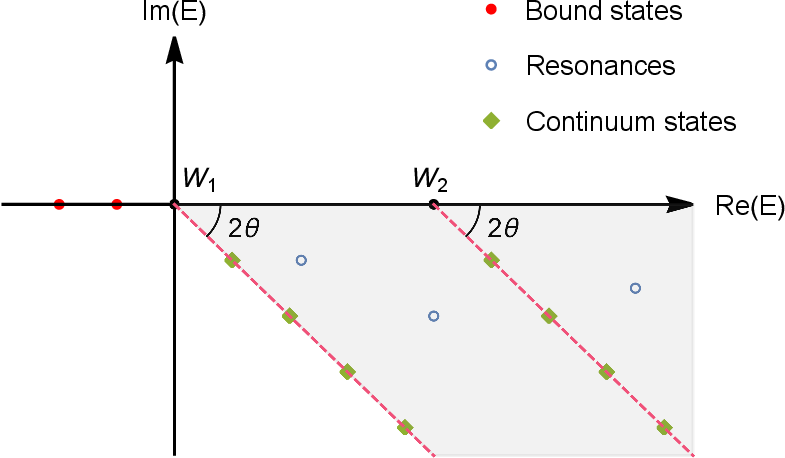}
    \caption{Schematic eigenvalue distributions of $H^{\theta}$ in the coupled-channel two-body systems.}
    \label{csmp}
\end{figure}
In the present work, the orbital wave functions are expanded in terms of a set of Gaussian basis functions. In the complex plane, the eigenvalues of the bound states and resonances are independent of the scaling angle $\theta$, while the continuum states change along with $\theta$. Then, a bound state or resonance can be pick out from the continuum states as shown in Figure~\ref{csmp}. 

Together with the masses, the realistic wave functions are also obtained. With the wave functions, one can calculate the root-mean-square (RMS) radii $r_{RMS}$ and component proportions $P$ for molecular states. For a resonance, the RMS radius $r_{RMS}$ and component proportion $P$ are defined by the c-product~\cite{T.yo,Lin:2023ihj,Shimizu:2016rrd}
\begin{eqnarray}
	r_{RMS}^2=(\psi^{\theta}|r^2|\psi^{\theta})=\sum_i\int r^2\psi^{\theta}_i(\bm{r})^2d^3\bm{r},\nonumber\\
	P=(\psi^{\theta}_i|\psi^{\theta}_i)=\int \psi^{\theta}_i(\bm{r})^2d^3\bm{r}\label{eq:c-prod},
\end{eqnarray}
where the $\psi^{\theta}_i$ is the wave function of the $i$-th channel and satisfies
the normalization condition
\begin{eqnarray}
	\sum_i(\psi^{\theta}_i|\psi^{\theta}_i)=1.
\end{eqnarray}
In this definition, the inner product is defined using the square of the wave function rather than the square of its modulus, which is the peculiar performance of a resonance. Moreover, it is worth emphasizing that the scaling angle $\theta$ should be larger than the $1/2Arg(\Gamma/2E)$ to ensure the normalizability of  wave functions for the resonant pole~\cite{Myo:2014ypa}.

\section{Results and discussions}\label{results}
\subsection{Isospin vector systems}
With the effective potentials, we can investigate the possible bound states and resonances by solving the coupled channel Schr\"odinger equation. In present work, only the cutoff $\Lambda$ reflecting the internal structures of interacting hadrons is a free parameter, which may be different for various coupled channel systems. Here, one can determine this model parameter by reproducing the masses of  $Z_b(10610)$ and $Z_b(10650)$ with the assumption that $Z_b(10610)$ and $Z_b(10650)$ are $I(J^{PC}) =1(1^{+-})$ molecular states. Considering the similarity of the relevant interactions resulted from the heavy quark symmetry, we prefer to use the same cutoff parameters for the $B^{(*)}_{(s)} \bar{B}^{(*)}_{(s)}$ system. In Ref.~\cite{Sun:2011uh}, considering single channel $B\bar{B}^*$ and $B^*\bar{B}^*$, the authors used parameter $g_s=g_\pi/(2\sqrt{6})=0.69$ to calculate the molecular states. In the present work, we apply the coupled-channel analysis on $B\bar{B}^*/B^*\bar{B}^*$ systems, where the coupled-channel effects is non-negligible. Moreover, the $g_s=2.82$ is used in our present calculations~\cite{Chen:2022dad,Wang:2019aoc,Machleidt:1987hj},  which can offer a stronger attractive force to form a molecule. Therefore, the cutoff value in our work is smaller than that of Ref.~\cite{Sun:2011uh}.  In   Ref.~\cite{Liu:2017mrh},  the authors performed a single channel analysis on $B\bar{B}^*$ and $B^*\bar{B}^*$ and they indicated that the $Z_{b}(10610)$ and $Z_b(10650)$ could not be molecules candidates. Also, they consider the $\eta-\eta^{\prime}$ mixing, which increases repulsive interaction for the isospin triplet. The different channels and interactions between Ref.~\cite{Liu:2017mrh} and ours give quite different results. We want to emphasis that the coupled-channel calculations are important for the $B^{(*)}_{(s)}\bar{B}^{(*)}_{(s)}$ systems.

The results of $I(J^{PC})=1(1^{+-})$ systems are shown in Table ~\ref{zb}. It can be seen that two weekly bound states are obtained in the $B\bar{B}^{*}/B^*\bar{B}^{*}$ coupled channel estimations within the range $\Lambda=950\sim1100~{\rm MeV}$. The first pole is just a few MeV below the $B\bar{B}^{*}$ threshold and dominated by the  $B\bar{B}^{*}+h.c.(^{3}S_{1})$ channel, which corresponds to the bottomonium-like state $Z_{b}(10610)$. For the second pole, it is dominated by $B^{*}\bar{B}^{*}$ channels and can be regarded as the $Z_{b}(10650)$. The $r_{RMS}$ of these two states are about $1\sim4 {\rm~ fm}$, which satisfy the sizes of hadronic molecules like the deuteron. Thus, we adopt the reasonable cutoff value in the range $800\sim 1200~{\rm MeV}$ to investigate other $B^{(*)}_{(s)}\bar{B}^{(*)}_{(s)}$ systems. Our results for all the systems depending on the cut off parameter $\Lambda$ are displayed in Figure~\ref{bb}.

\begin{table*}[htbp]
	\renewcommand\arraystretch{1.4}
	\caption{\label{zb} The molecular states for $I(J^{PC})=1(1^{+-}) $ system. The numbers in the bracket are the components proportion of $B \bar{B}^\ast+h.c.$ and  $B^\ast \bar{B}^\ast$channels.}
	\begin{ruledtabular}
		\begin{tabular}{ccccccccc}
			&$\Lambda(\rm{MeV})$
			&$r_{RMS}(\rm{fm})$
			&$E(\rm{MeV})$&($B\bar{B}^{*}(^{3}S_{1})$&$B\bar{B}^{*}(^{3}D_{1})$&$B^{*}\bar{B}^{*}(^{3}S_{1})$&$B^{*}\bar{B}^{*}(^{3}D_{1})$&$B^{*}\bar{B}^{*}(^{5}D_{1})$)
			\\\hline
			&\multirow{2}{*}{950} &2.75&10604& (98.91&0.61&0.00&0.00&0.47)\\
			&&4.22&10649& (0.00&0.00&99.65&0.35&0.00)\\
			&\multirow{2}{*}{1000}&1.44&10603&(98.38&0.81&0.00&0.00&0.81) \\
			&&2.67&10649&(0.00&0.00&99.45&0.55&0.00) \\
			&\multirow{2}{*}{1050} &1.39&10601&(97.91&0.94&0.00&0.00&1.14) \\
			&&1.84&10648&(0.00&0.00&99.31&0.69&0.00) \\
			&\multirow{2}{*}{1100} &1.14&10599&(97.46&1.05&0.00&0.00&1.48) \\
			&&1.43&10646&(0.00&0.00&99.22&0.78&0.00) \\
		\end{tabular}
	\end{ruledtabular}
\end{table*}

Besides $Z_b(10610)$ and $Z_b(10650)$ states, we also have $1(0^{++})$ $B\bar{B}/B^*\bar{B}^*$, $1(1^{++})$ $B\bar{B}^{*}/B^*\bar{B}^*$, and $1(2^{++})$ $B^*\bar{B}^*$ coupled channels for the isospin vector systems. 
As for $I(J^{PC})=1(0^{++})$ system, a bound solution around 10557 MeV is obtained and the corresponding RMS radius is about $1.56\sim 4.02~{\rm fm}$.  For the $I(J^{PC})=1(1^{++})$ system, the mass of obtained bound state varies from 10591 to 10604 MeV when the cutoff $\Lambda$ increases from 950 to 1100 MeV, which is a possible molecular candidate. Finally, for the $I(J^{PC})=1(2^{++})$ system, when the cut off value lie in $800\sim 1200 {\rm MeV}$, one can not obtain any structure in the complex plane. If we increase the cutoff $\Lambda$ even larger, a possible bound solution appear. The reason is that the $1(2^{++})$ system only have a $B^*\bar{B}^*$ channel, and the attraction is somewhat weaker than the $1(0^{++})$ and $1(1^{++})$ systems.

\subsection{Isospin scalar systems}

\begin{figure*}
	\centering
	\includegraphics[scale=0.60]{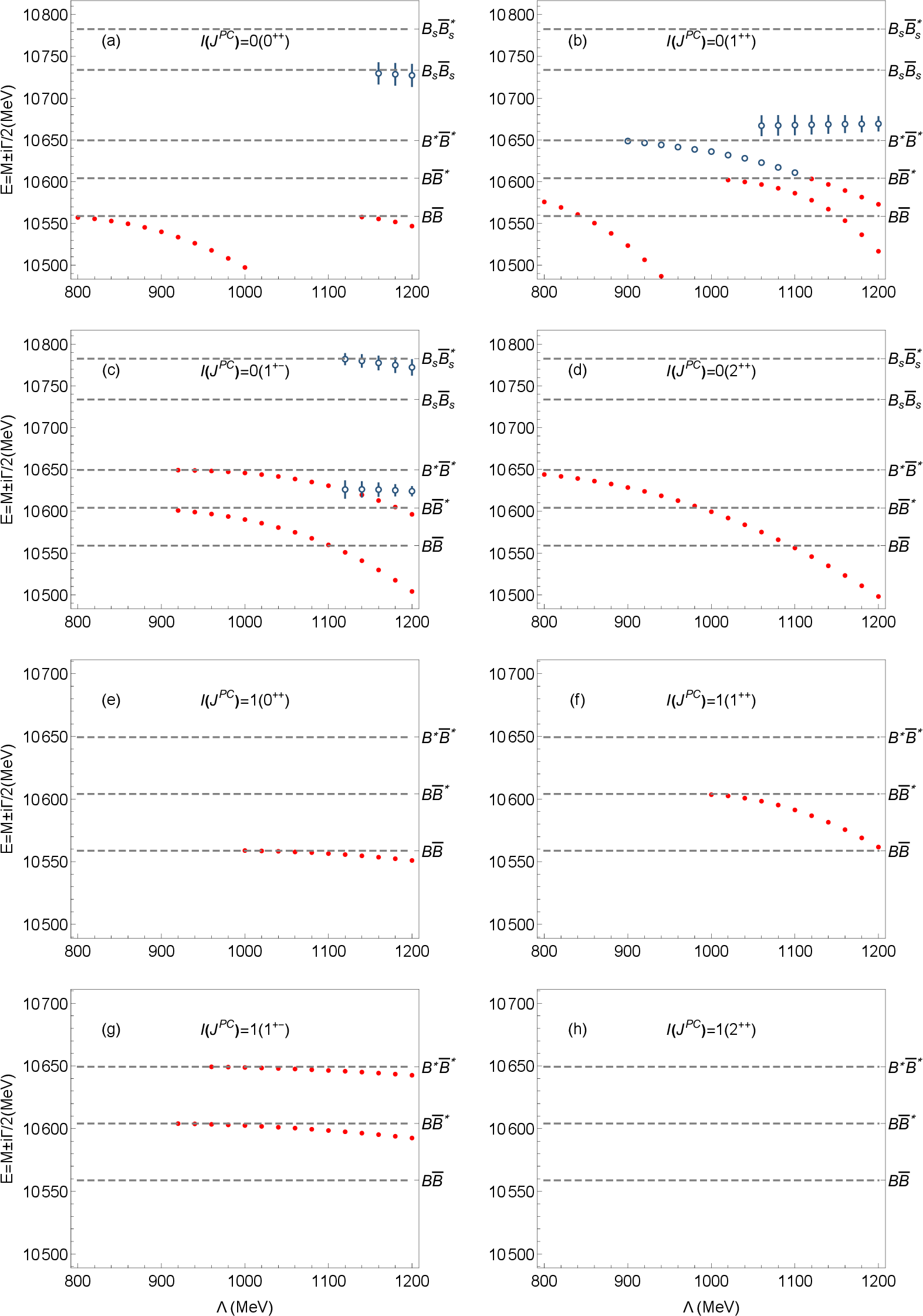}
	\caption{The $\Lambda$  dependence for the $B^{(*)}_{(s)}\bar{B}^{(*)}_{(s)}$ systems. The red solid dots stand for the bound states. The blue open circles with bars correspond to the resonances, and the bars represent the total widths for resonances.}
	\label{bb}
\end{figure*}

Unlike the isovector systems, more coupled channels arising form bottom-strange mesons are involved in the $I=0$ systems. We first deal with the $I(J^{PC})=0(0^{++})$ $B\bar{B}/B_s\bar{B}_s/B^*\bar{B}^*/B_s^*\bar{B}_s^*$  channels. From Figure~\ref{bb}, a bound state emerges with $\Lambda=800~\rm{MeV}$, and becomes deeper when the cutoff $\Lambda$ increases. With the deep binding energy, the $r_{R.M.S}$ radius of this state is also very small. Since the $r_{RMS}$ of $B$ and $B^*$ mesons are predicted to be 0.42 and 0.45 fm respectively~\cite{Godfrey:1985xj,Godfrey:2016nwn}, one expect that the sizes for $B\bar{B}^{*}/B^*\bar{B}^{*}$ molecular states should be larger than 0.9 fm. Then, we prefer to eliminate the obtained bound states or resonances with sizes less than 0.9 fm, which can not be regarded as molecular states. Increasing the $\Lambda$ near 1200 MeV, we obtain a bound and resonate states below the $B\bar B$ and $B_s\bar {B}_s$ thresholds, respectively. Although the   
$r_{RMS}$ radii of these two states support the molecular interpretation, the cutoff seems a bit larger than that of the $Z_b(10610)$ and $Z_b(10650)$ states.

\begin{figure*}[!htbp]
	\centering
	\includegraphics[scale=0.6]{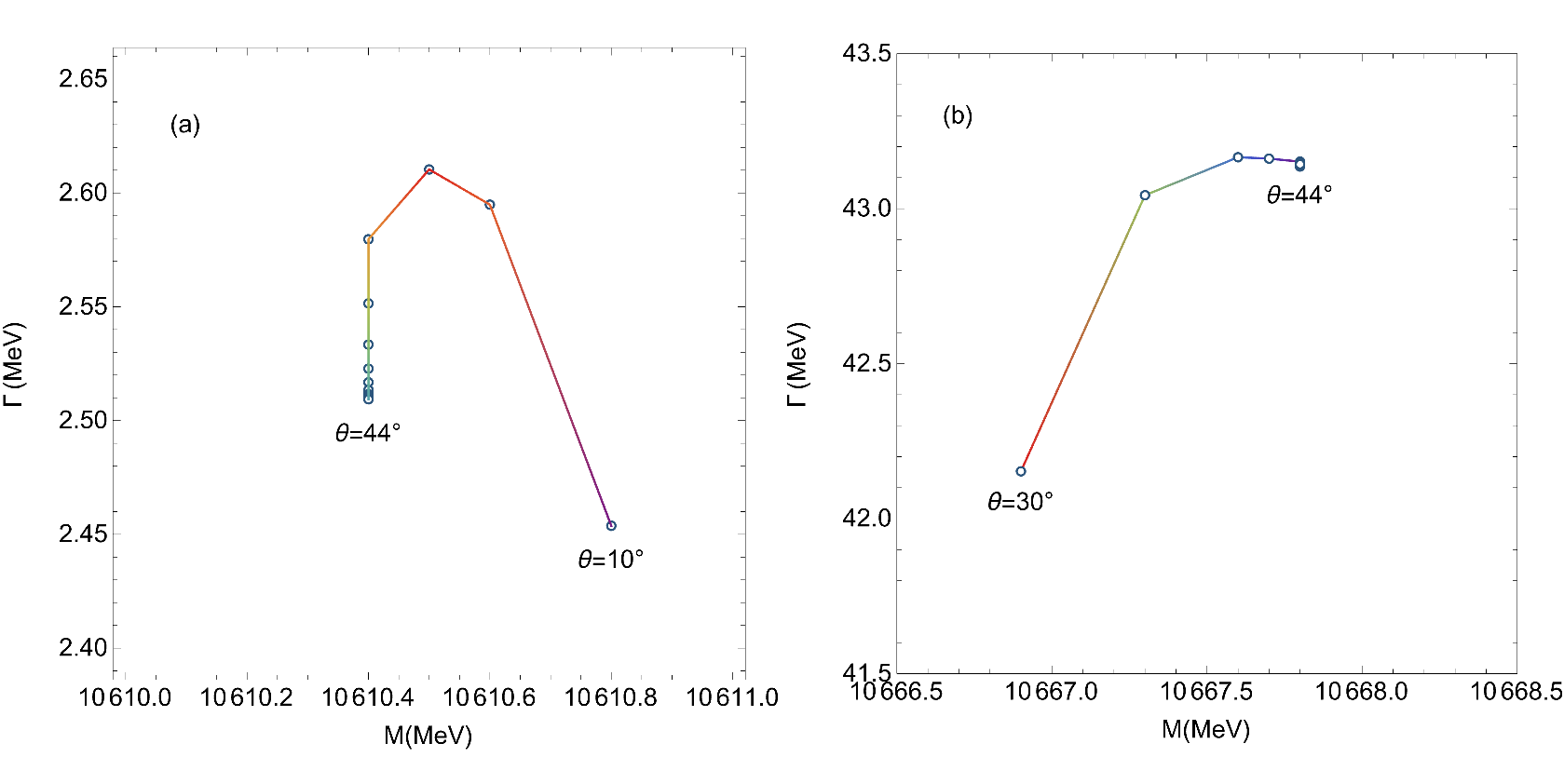}
	\caption{\label{re} The variations of complex energies for the two predicted $0(1)^{++}$ resonances with $\theta$.}
\end{figure*}

For the  $0(1^{++})$ $B\bar{B}^{*}/B^{*}\bar{B}^*/B_s\bar{B}_{s}^{*}/B_s^*\bar{B}_s^*$ system, a similar deep bound state as $0(0^{++})$ exist, which is also ignored here.  When the cutoff lie in the range of $\Lambda=950\sim1100~{\rm MeV}$, one can find a bound state dominated by the $B\bar B^*$ channel, and two resonances above $B\bar B^*$ and $B^*\bar B^*$ thresholds respectively. Interestingly, one can see the narrow resonance turn into a bound state if the attractive interaction increases. Also, we present the variations of complex energies for the two  predicted $0(1)^{++}$ resonances with $\theta$ in Figure~\ref{re}, where one can find that the numerical errors caused by the $\theta$ are tiny and rapidly converge with increasing $\theta$. For the $I(J^{PC})=0(1^{+-})$ system, two bound states below the $B\bar B^*+h.c.$ and $B^*\bar B^*$ thresholds emerge, which can be regarded as exotic $0(1^{+-})$ molecular states. These two bound states are isoscalar cousins of $Z_b(10610)$ and $Z_b(10650)$ particles, which can be searched for in future. Moreover, the differences between $0(1^{++})$ and $0(1^{+-})$ is caused by the $C-$parity, which reflects the different flavor wave functions and interference.

For the  $I(J^{PC})=0(2^{++})$ $B^{*}\bar{B}^*/B_s^*\bar{B}_s^*$ system, only one bound state are obtained. With the binding energy increasing, the RMS radius becomes too compact in a reasonable range of cutoff parameter $\Lambda$. Also, our results indicate that the attraction of $0(2^{++})$ system is weaker than the $0(0^{++})$ and $0(1^{++})$ systems, which is similar with the $I=1$ cases. Moreover, when the cutoff $\Lambda$ belongs to $950\sim1100$ MeV, although the $B_s^{(*)}\bar{B}_s^{(*)}$ channels are included in the coupled-channel calculations, no molecular state is found near $B_s^{(*)}\bar{B}_s^{(*)}$ thresholds in present work.

\subsection{Further discussions}

 We collect the final results of  $B^{(*)}_{(s)}\bar{B}^{(*)}_{(s)}$ systems in Table~\ref{sum}. Our results show that the bound states or resonances appear in the $J=0$ and 1 systems, while the formation of molecules in $J=2$ systems is harder. Meanwhile, both bound states and resonances are found in the $I=0$ systems, while only several bound states exist in the $I=1$ systems. That is to say, the systems with higher isopsipn or spin are harder to form molecular states than the lower ones. In present work, the $J=2$ or $I=1$ case has less channels compared with others, which suggests that the coupled channels may play crucial roles to obtain these molecules. Moreover, the dependence on spin and isospin for potentials is also reflected by the Clebsch-Gordan coefficients, which determines the relative sign and strength.
 
    \begin{table*}[!htbp]
 	\renewcommand\arraystretch{1.4}
 	\caption{\label{sum} The summary of our predictions for $B^{(*)}_{(s)}\bar{B}^{(*)}_{(s)}$ systems with cutoff $\Lambda$ in a range of $950\sim1100$ $\rm MeV$. Here, the $"-"$ means nonexistence, $"\checkmark"("\times")$ represents that the corresponding state may (may not) form a molecular state.}
 	\begin{ruledtabular}
 		\begin{tabular}{cccccccc}
 			&$I(J^{PC})$
 			&Mass$(\rm MeV)$&Width$(\rm MeV)$&$r_{RMS}(\rm fm)$&Status& Selected decay mode
 			\\\hline
 			&$0(0^{++})$&$10522\sim10424$&$-$&$0.63\sim0.46$&$\times$&$-$\\\hline
 			&\multirow{4}{*}{$0(1^{++})$}&$10476\sim10194$&$-$&$0.36\sim0.24$&$\times$&$-$\\&&$10602\sim10586$&$-$&$1.85\sim0.91$&$\checkmark$&$\eta_b(1S)\omega/\Upsilon(nS)\eta/\Upsilon(1S)\eta^\prime$  \\&&$10643\sim10610$&$9.72\sim2.50$&$1.30+0.37i\sim0.85+0.49i$&$\checkmark$ &$\eta_b(1S)\omega/\Upsilon(nS)\eta/\Upsilon(1S)\eta^\prime/B\bar B^*+h.c.$ \\
 			&&$10667\sim10668$&$46.04\sim43.96$&$3.28+1.89i\sim2.94+1.69i$&$\checkmark$&$\eta_b(1S)\omega/\Upsilon(nS)\eta/\Upsilon(1S)\eta^\prime/B\bar B^*+h.c./B^*\bar B^*$ \\\hline	
 			&\multirow{2}{*}{$0(1^{+-})$}&$10598\sim10560$&$-$&$1.32\sim0.70$&$\checkmark$ &$\eta_b(1S)\omega/\Upsilon(nS)\eta/\Upsilon(1S)\eta^\prime$ \\
 			&&$10649\sim10631$&$-$&$2.67\sim0.90$&$\checkmark$&$\eta_b(1S)\omega/\Upsilon(nS)\eta/\Upsilon(1S)\eta^\prime$  \\\hline	
 			&$0(2^{++})$&$10616\sim10556$&$-$&$0.65\sim0.48$&$\times$&$-$\\\hline
 			&$1(0^{++})$&$10559\sim10557$&$-$&$4.02\sim1.56$&$\checkmark$&$\eta_b(nS)\pi$/$\Upsilon(1S)\rho$ \\\hline
 			&$1(1^{++})$
 			&$10604\sim10591$&$-$&$2.56\sim0.70$&$\checkmark$  &$\eta_b(1S)\rho/\Upsilon(nS)\pi$ \\\hline
 			&\multirow{2}{*}{$1(1^{+-})$}&$10604\sim10599$&$-$&2.75$\sim$1.14&$\checkmark$ &$\eta_b(1S)\rho/\Upsilon(nS)\pi$ \\
 			&&$10649\sim10646$&$-$&4.22$\sim$1.44&$\checkmark$&$\eta_b(1S)\rho/\Upsilon(nS)\pi$   \\\hline
 			&$1(2^{++})$&$-$&$-$&$-$&$\times$&$-$ \\
 		\end{tabular}
 	\end{ruledtabular}
 \end{table*}

The importance of coupled channels effects is reflected in the formation of those bound or resonant states in Table~\ref{sum}. For the $I=1$ case, since the $\omega$ and $\sigma$ exchange potentials are strong enough, the single $1(0^{++})B\bar{B}$ channel is enough to form a loosely bound state. However, if only single $1(1^{++})B\bar{B}^{*}$ channel is considered, one cannot find any pole even with a large cutoff parameter, which demonstrates that the couplings with $B\bar B$ and $B^*\bar B^*$ channels are indispensable to form the  $I(J^{PC})=1(1^{++})$ molecular state. For the $I(J^{PC})=0(1^{++})$ case, it can be found that the long-range pion exchange potential is crucial to form the resonance, the $\rho$ and $\omega$ meson exchange interactions will affect the widths of resonances, and the $\eta$ exchange potential is negligible. For the $I(J^{PC})=0(1^{+-})$ case, when the single channel $B\bar{B}^*$ and $B^*\bar{B}^*$ are calculated by using one pion exchange separately, they can form the shallow bound states, which  is also found in Ref.~\cite{Sun:2011uh}.  Then, the coupled channel interactions deepen the binding energies of two $0(1^{+-})$ bound
states as the ispospin partners of the $Z_b(10610)$ and $Z_b(10650)$.

It is also interesting to note that the root mean square $r_{RMS}$ for the predicted $0(1^{++})$ resonances are complex numbers. Under the circumstances, one can choose an interpretation scheme proposed by T. Berggren, where the definition of the expectation value is generalized from bound state to resonance ~\cite{Berggren:1970wto}. That is to say, the real part stands for the ordinary physical expectation value and the imaginary part corresponds to a measure of the uncertainty in observation. The numerical calculations of $r^2$ supported this generalized interpretation~\cite{Gyarmati:1972yac, 1997matrix}.

Besides the experimentally observed  $Z_b(10610)$ and $Z_b(10650)$, we also obtain some possible molecular states in the $0(1^{+\pm})$, $1(0^{++})$, and $1(1^{++})$ systems.  Analogous to the charmonium-like states $X(3872)$ and $X(4014)$, these predicted $X_b$ and $Z_b$  states are most likely to discovered in experiments. From Review of Particle Physics~\cite{ParticleDataGroup:2022pth}, the conventional $3P$ bottomium are around 10513 MeV, and the predicted $b\bar b(4P)$ states in relativized quark model are about $10775 \sim 10798$ MeV~\cite{Godfrey:2015dia}. It can be seen that our predicted molecular states subtly keep away from the conventional bottomium. Hence, the observation of new hadrons in this energy region immediately demonstrate the exotic nature. According the masses and quantum numbers, we present some possible decay modes of these predicted states in Table~\ref{sum}. It can be seen that the $\eta_b(nS)/\Upsilon(nS)$ plus light mesons are the excellent final states to search for the bound states, while the $B\bar B^*+h.c.$ and $B^* \bar B^*$ channels are suitable for the resonances. We highly recommend that the LHCb and Belle II Collaborations can hunt for these bottom exotic particles in future.

\section{Summary}\label{sec4}

In present work, we study the $B^{(*)}_{(s)}\bar{B}^{(*)}_{(s)}$ systems in a coupled-channel approach with the one-boson-exchange potentials. The coupled-channel Schr\"odinger equation are solved by using the Guassian expansion method and complex scaling method. Firstly, we analyze the $I(J^{PC})=1(1^{+-})$ $B\bar{B}^{*}/B^{*}\bar{B}^{*}$ system to describe the $Z_{b}(10610)$ and $Z_{b}(10650)$ particles as molecular states, which help us determine the reasonable range of cutoff parameter $\Lambda$. Then, we systematically study various $B^{(*)}_{(s)}\bar{B}^{(*)}_{(s)}$ combinations with different quantum numbers to search for possible bound states and resonances. Some bound states and resonances appear in the isoscalar systems, while only several shallow bound states exist for isovector systems. Also, our results indicate that the systems with higher isopsipn or spin are harder to form molecular states than the lower ones.

With $\Lambda=950\sim1100~\rm{MeV }$, all the predicted exotic states locate near the $B\bar{B}$, $B\bar{B}^{*}+h.c.$, and $B^{*}\bar{B}^{*}$ thresholds, while no molecule is found near the $B_s^{(*)}\bar{B}_s^{(*)}$ thresholds in present work. These states named as $X_b$ or $Z_b$ can be regarded as the partners of charmonium-like states $X(3872)$, $X(4014)$, $Z_c(3900)$, and $Z_c(4020)$. Their masses lie far away from the excited conventional $P-$wave bottomium, which can be easily identified both theoretically and experimentally. Moreover, the $\eta_b(nS)/\Upsilon(nS)$ plus light mesons are the excellent final states to search for the bound states, while the $B\bar B^*+h.c.$ and $B^* \bar B^*$ channels are suitable for the resonances. We hope our present calculations can provide helpful information, and highly recommend that the future experiments can search for these bottom exotic particles.

\subsection*{ACKNOWLEDGMENTS}
We would like to thank Rui Chen, Xian-Hui Zhong, Fu-Lai Wang, and Mu-Yang Chen for useful discussions. This work is supported by the Natural Science Foundation of Hunan Province under Grant No. 2023JJ40421 and the Key Project of Hunan Provincial Education Department under Grant No. 21A0039. This work is also supported by the National Natural Science Foundation of China under Grant No. 12175037, No. 12335001, and No. 12375142, the National Key Research and Development Program of China under Contract No. 2020YFA0406300, and the Sino-German CRC 110 Symmetries and the Emergence of Structure in QCD project by National Natural Science Foundation of China under Grant No. 12070131001.

\end{document}